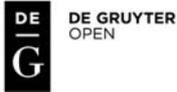

ORIGINAL PAPER

# Application of instruments of nuclear physics to the calculation of theoretical dose distributions in various organs of the human body for beams used in hadrontherapy

Weronika Maliszewska,
Przemysław Sękowski,
Izabela Skwira-Chalot

**Abstract.** The area of interests of nuclear physics are studies of reactions, wherein atomic nuclei of projectile collide with target nuclei. An amount of energy lost by projectile nucleus during its passing through the target is a major issue – it is important to know how charged particles interact with matter. It is possible to afford this knowledge by using theoretical programs that calculate energy loss applying the Bethe-Bloch equation. Hadrontherapy, which is a field of still growing interest, is based on the interactions of charged particles with matter. Therefore, there exists a need of creating a simple model that could be used to the calculation of dose distributions in biological matter. Two programs (SRIM, Xeloss), used to the calculation of energy loss by nuclear physicist, have been adapted to determine the dose distributions in analogues of human tissues. Results of the calculations with those programs for beams used in hadrontherapy (e.g. $^1$H, $^{12}$C) will be compared with experimental data available in references.

W. Maliszewska✉, P. Sękowski, I. Skwira-Chalot
Faculty of Physics,
University of Warsaw,
5 Pasteura Str., 02-093 Warsaw, Poland,
E-mail: wmaliszewska@student.uw.edu.pl

## Introduction

Hadrontherapy, because of the radiobiological advantages (depth to dose distribution, reduction of radiation dose in patients' body, smaller sensitivity for oxygen-depleted tissues), is very often used in tumor treatment [1]. High effectiveness of this kind of therapy decreases probability of recurrence of tumor [2, 3]. For efficient tumor treatment, we need to know two things: how the charged particles interact with biological matter and how big amount of energy is deposited during passing into tissues. These two 'puzzles' give an information about a dose that patient received during therapy. Because experiments on animals or humans tissues are rather complicated, we need to create a simple model (or theoretical program) that could be used to calculate the dose distributions in biological matter. For this purpose, two programs were tested: SRIM [4] and Xeloss, which are usually used in the calculation of energy loss by nuclear physicist.

## Theoretical calculations

SRIM is a program that calculates stopping power and range of ion in matter using Monte Carlo simulation method named binary collision approximation (BCA). Program Xeloss calculates stopping power and range by applying the Bethe-Bloch formula [5, 6]:

$$(1) \quad -\frac{dE}{dx} = \frac{4\pi}{m_e c^2} \cdot \frac{N_A \rho Z z^2}{A M \beta^2} \cdot \left(\frac{e^2}{4\pi\varepsilon_0}\right)^2 \cdot \left[\ln\left(\frac{2 m_e c^2 \beta^2}{\langle I \rangle (1-\beta^2)}\right) - \beta^2\right]$$

This equation describes the energy loss per distance travelled of swift charged particle with speed $v$ charge $z$, energy $E$, travelling a distance $x$ into a matter with mean excitation potential $\langle I \rangle$; $c$ is the speed of light; $\varepsilon_0$ the vacuum permittivity; $\beta = v/c$, $e$ and $m_e$ the electron charge and mass; $\rho$ is the density of the material; $Z$ its atomic number; $A$ its relative atomic mass; $N_A$ the Avogadro number and $M$ the molar mass constant.

### SRIM or Xeloss? Differences between programs

The programs, which we tested, calculate range of ions in matter by applying different algorithms. Therefore, we can check what is a difference in range calculations between SRIM and Xeloss. Figure 1 presents difference of ranges in water for ion beams of hydrogen, carbon and oxygen in the range of energy from 200 to 400 MeV/nucleon. We can see that for a given beam, the difference in range calculations increases with energy. Moreover, with increasing atomic number of beam, the differences in range calculations decrease. Those differences are significant and should be taken into account in further calculations.

### Ions in water

Before the application of ion beam in therapy of tumors, it is important to do some experimental measurements of depth-dose distribution (DDD) in water phantoms, which approximate very well soft tissues. In the first step of our analysis, the Bragg curves for protons of energy of 60 MeV/nucleon were calculated. Figure 2 presents the dose distribution, which was calculated by normalization of deposited energy in water to the maximum value (at Bragg peak). (All DDDs curves were obtained by the same procedure). One can see that the range of protons in water calculated with SRIM reconstructs the experimental data better than Xeloss calculations. The position of Bragg peak was obtained with 6% and 24% accuracy for SRIM and Xeloss, respectively. The reason for these differences may be the problem with estimation of mean excitation potential $\langle I \rangle$ (location of Bragg peak is strongly dependent from mean excitation potential [8]). Uncertainties were calculated using the formula

$$(2) \quad \varepsilon = \left|\frac{x_m - x_t}{x_t} \cdot 100\%\right|$$

where $x_m$, $x_t$ are the values of experimental and theoretical ranges, respectively.

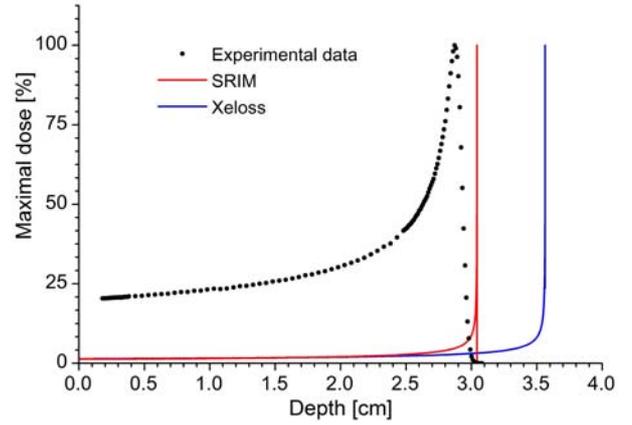

**Fig. 2.** Depth-dose distribution of protons in water at beam energy of 60 MeV/nucleon. Experimental data thanks to J. Swakoń from CCB IFJ PAN.

The next step of our analysis was devoted to the energy loss of oxygen ion beams in water. Figure 3

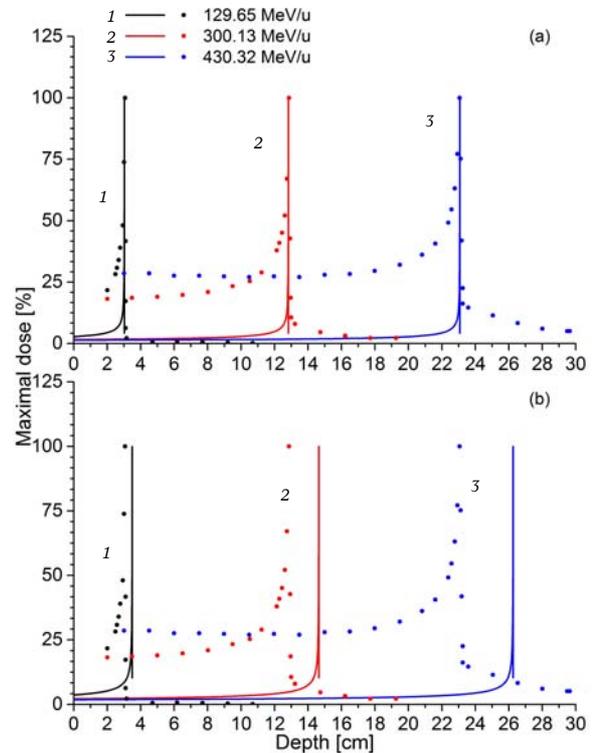

**Fig. 3.** DDD of oxygen ions in water for three various energies calculated with (a) Xeloss and (b) SRIM. Dots and lines represent the experimental and theoretical data, respectively. (Uncertainties of experimental data are not available in paper [7]).

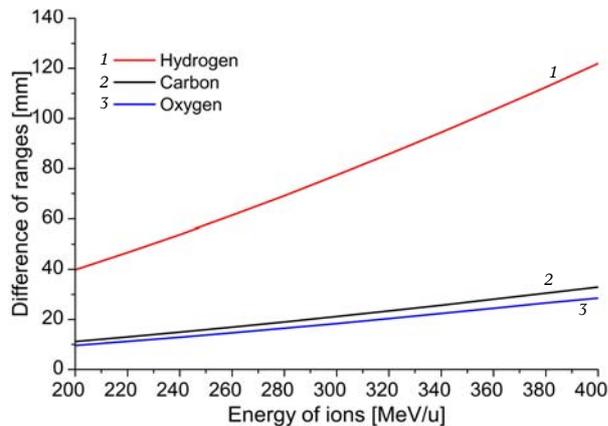

**Fig. 1.** Difference between ranges of oxygen, carbon and hydrogen ions in water calculated with SRIM and Xeloss.

presents comparison of theoretical DDD with experimental ones [7]. The theoretical and experimental results were obtained for three different values of beam energies: 129.65, 300.13 and 430.32 MeV/nucleon. One can see that for each value of energy, SRIM (Fig. 3b) reconstructed the experimental range quite well. The Bragg peak positions are determined with average accuracy at the level of 0.5% in case of SRIM and 14% for Xeloss.

**Carbon ions in tissues**

To compare our calculations with experimental data for carbon ions in different tissues [8], for both programs separately, energy of the carbon beam was chosen in order to reconstruct experimental position of the Bragg peak for water. This procedure allows to estimate value of energy of carbon ions after passing through initial elements of experimental system and enables studying precision of programs for ranges of ions in each tissues. Obtained energies for carbon ions were equal to 280 MeV/nucleon/260.9 MeV/nucleon for calculation of SRIM/Xeloss. Dose distributions measured [9] and calculated are presented in Fig. 4. It has been assumed that pig tissues have composition similar to that of human tissues. Paper [10], which characterizes the composition of the human body, does not determine the composition of brain; therefore, for the purposes of calculations, it was assumed that brain consists of 10% of cerebrospinal fluid, 40% of grey matter and 50% of white matter.

The theoretical position of Bragg peak calculated with SRIM (Fig. 4b) are close to peaks from experimental data, except for adipose tissue (too big range) and kidney tissue (too small range). Differences between experimental and theoretical data may be caused by improper approximation of mean excitation potential $<I>$. Also composition of human tissues differs from pig tissues and this factor may have influence on results of theoretical depth too. Uncertainty of calculated depth where carbon ions stop in tissue equals for SRIM 0.68%. Figure 4a presents calculation results obtained with Xeloss. The graph shows that Xeloss, in contrast to SRIM, correctly recreates the range of carbon ions in adipose tissue but unfortunately gives bigger (than SRIM) differences between theoretical and experimental data in Bragg peaks positions for liver, kidney and brain. The reason of these differences could be incorrect estimation of mean excitation potential $<I>$. Uncertainty for Xeloss calculations equals 0.95%.

**Conclusions**

Presented analysis allows to say that by the application of simple programs, it is possible to correctly define the ranges of ions in various tissues. On this basis, it is also possible to predict dose distributions with a good approximation, except dose distribution in water for ions in water calculated with Xeloss. Uncertainties of determining ranges on the level of 1% in tissues for both programs confirm that SRIM and Xeloss may be used to calculate a dose and a ranges at the beginning of planning experiments or hadrontherapy.

**References**

1. Particle Therapy Co-Operative Group. (2014). *Patient statistics per end of 2013*. Retrieved August 15, 2014, from http://www.ptcog.ch/archive/patient_statistics/.
2. Durante, M., & Loeffler, J. (2010). Charged particles in radiation oncology. *Nat. Rev. Clin. Oncol.*, 7(1), 37–43. DOI: 10.1038/nrclinonc.2009.183.
3. Schulz-Ertner, D., Jäkel, O., & Schlegel, W. (2006). Radiation therapy with charged particles. *Semin. Radiat. Oncol.*, 16(4), 249–259. DOI: 10.1016/j.semradonc.2006.04.008.
4. Ziegler, J. (2013). SRIM [computer software].
5. Bloch, F. (1933). Bremsvermögen von Atomen mit mehreren elektronen. *Z. Phys. A: Hadrons Nucl.*, 81(5/6), 363–376. DOI: 10.1007/BF01344553.
6. Bloch, F. (1933). Zur bremsung rash bewegter teilchen beim durchgang durh materie. *Ann. Phys.*, 408(3), 285–320. DOI: 10.1002/andp.19334080303.
7. Kurz, C., Mairani, A., & Parodi, K. (2012). First experimental-based characterization of oxygen ion beam depth dose distributions at the Heidelberg Ion-Beam Therapy Center. *Phys. Med. Biol.*, 57(15), 5017–5034. DOI: 10.1088/0031-9155/57/15/5017.
8. Soltani-Nabipour, J., Sardari, D., & Cata-Danil, G. H. (2009). Sensitivity of the Bragg peak curve to the average ionization potential of the stopping medium. *Rom. J. Phys.*, 54(3/4), 321–330.

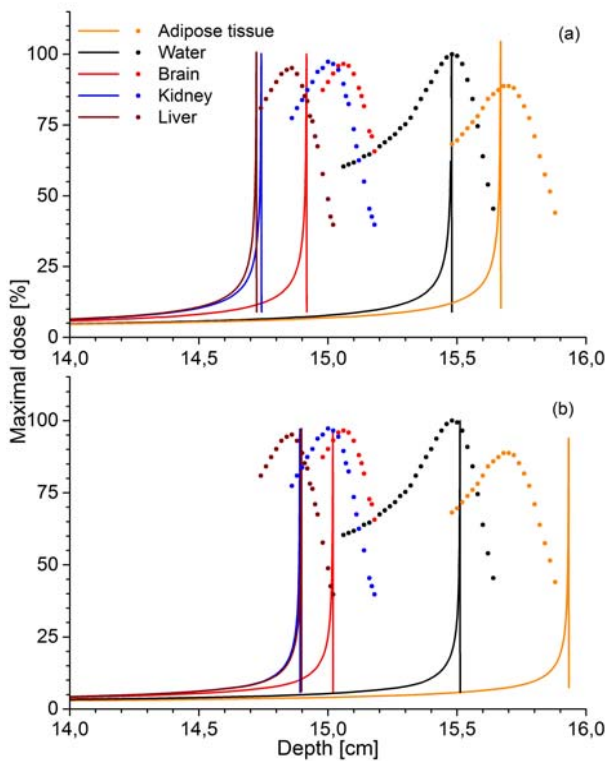

**Fig. 4.** Carbon ions in various tissues for energy (a) 260.9 MeV/u calculated with Xeloss and (b) 280 MeV/u calculated with SRIM. Dots and lines represent the experimental and theoretical data, respectively. (Uncertainties of experimental data are not available in paper [9]).

9. Rietzel, E., Schardt, D., & Haberer, T. (2007). Range accuracy in carbon ion treatment planning based on CT-calibration with real tissue samples. *Radiat. Oncol.*, *2*, 14(9 pp.). DOI: 10.1186/1748–717X-2-14.

10. Woodard, H., & White, D. (1986). The composition of body tissue. *Br. J. Radiol.*, *59*(708), 1209–1219.